\documentclass[10pt,conference]{IEEEtran}
\pdfoutput=1
\IEEEoverridecommandlockouts
\usepackage{amsmath,amssymb,amsfonts}
\usepackage[utf8]{inputenc}
\usepackage{algpseudocode}
\usepackage{cite}
\usepackage{algorithm}
\usepackage{graphicx}
\usepackage{textcomp}
\usepackage{url}
\usepackage[caption=false]{subfig}
\usepackage[justification=centering]{caption}
\usepackage{epsfig}
\usepackage{float}
\usepackage{color}
\usepackage{balance}
\usepackage{bbm}
\usepackage{textcomp}
\usepackage{enumitem}
\usepackage{mathtools}
\ifCLASSINFOpdf
\else
\fi
\begin{document}

\title{ Blue Communications for Edge
Computing: the  Reconfigurable Intelligent Surfaces Opportunity}

\author{Fatima Ezzahra Airod, Mattia Merluzzi, Antonio Clemente, Emilio Calvanese Strinati\\
CEA-Leti, Université Grenoble Alpes, F-38000 Grenoble, France\\

email:\{fatima-ezzahra.airod, mattia.merluzzi, antonio.clemente, emilio.calvanese-strinati\}@cea.fr.\thanks{The work of F. Ezzahra Airod, A. Clemente, and E. Calvanese Strinati was supported by the EU H2020 RISE-6G project under grant number 101017011.}\vspace{-.4 cm}}

\maketitle

\begin{abstract}
Wireless traffic is exploding, due to the myriad of new connections and the exchange of capillary data at the edge of the networks to operate real-time processing and decision making. The latter especially affects the uplink traffic, which will grow in 6G and beyond networks, calling for new optimization metrics that include energy, service delay, and electromagnetic field (EMF) exposure (EMFE). To this end, reconfigurable intelligent surfaces (RISs) represent a promising solution to mitigate the EMFE, thanks to their ability of shaping and manipulating the impinging electromagnetic waves. In line with this vision, this paper proposes an online adaptive method to mitigate the EMFE under end-to-end delay constraints of a computation offloading service, in the context of RIS and multi-access edge computing (MEC)-aided wireless networks. 
The goal is to minimize the long-term average of the EMF human exposure under such constraints, investigating the advantages of RISs towards \textit{blue} (i.e. low EMFE) communications. A multiple-input multiple-output (MIMO) system  is investigated as part of the visions towards 6G.  
Focusing on a typical scenario of computation offloading, the method jointly  and adaptively optimizes user precoding, transmit power,  RIS reflectivity parameters, and receiver combiner, with theoretical guarantees on the desired long-term performance. Besides the theoretical results, numerical simulations assess the performance of the proposed algorithm, when exploiting accurate antenna patterns, thus showing the advantage of the RIS and that of our method, compared to benchmark solutions.
\end{abstract}

\vspace{0.1 cm}

%
\IEEEpeerreviewmaketitle
\section{Introduction}
The research on 6G has revolutionary long term ambitions, among which, building upon 5G, the millimeter wave (mmWave) and even sub-teraHertz (THz) bands represent key drivers, to cope with the explosion of data volumes communicated through smart and fully reconfigurable networks. Indeed, this is undeniably marked due to an unprecedented revolution of applications, such as immersive virtual reality, connected autonomous systems, the industrial internet of things, and many other vertical sectors \cite{1,2}. More precisely, as  predicted by the international telecommunication union (ITU), this exponential growth in data is expected to reach 5 zettabytes (ZB) per month by 2030 \cite{6G}. This explosion pertains to also uplink communications \cite{UL}, due to the continuous transfer of capillary data from extreme edge devices such as sensors and cars, to enable computational demanding services in real-time, at the edge of wireless communication networks. A key enabler of this vision is undoubtly the push of storage and computing at the wireless edge, thanks to the arising paradigm of multi-access edge computing (MEC) \cite{MEC}.
To be effective, these services require to be enabled with new levels of dependability, reliability and sustainability. From a radio access perspective,  the mmWave bands (30 GHz to 300 GHz) represent a good fit to face the aforementioned challenges with higher offered data rates (Gbps). However, these communications generally suffer from high path loss and unexpected blockages \cite{blc1}. To this end, beamforming (BF) is used to enable highly directional communications. However, in some cases, this may intensify electromagnetic field (EMF) exposure (EMFE), which has been extensively shown to be a myth \cite{Chiaraviglio_myth,emf_org}.  Hence, it is fundamental to design mitigation techniques for a sustainable development and a smooth public acceptance of such new technologies \cite{MerluzziEMF}. To this end, reconfigurable intelligent surfaces (RISs), have recently emerged as a promising candidate to counteract the above mentioned issues and others, thanks to their ability to
shape and manipulate the impinging electromagnetic waves.
More specifically, in the reflective case, RISs are composed of radiating elements that can be adaptively configured to shape the reflection of the incident wave. Therefore, owing to their abilities to customize time-varying wireless propagation environments, RISs are definitely a key ingredient of future networks, due to the fact that short and long-term requirements can be dynamically controlled in specific locations in space and time \cite{6,7}.\\ 
\textbf{Related works.} In the recent literature, a few  RIS-based solutions have been investigated to address resource allocation schemes, under EMF exposure constraints \cite{emf_org,8,9}. Such contributions tackle typical communication problems such as spectral efficiency maximization or energy consumption minimization. In \cite{emf_org}, both RIS and massive MIMO BF have been considered and jointly configured to build a spectrally and energy efficient radio link under an EMFE constraint. In \cite{8}, the authors configure RIS phases to minimize the population EMFE subject to quality of service constraints. Then, in \cite{9}, the energy efficiency maximization has been investigated under both power and EMF constraints. However, to the best of our knowledge,  none of these works  has focused on boosting trade-offs related to EMF human exposure and computation offloading service requirements, in MIMO and RIS-aided wireless systems. A first attempt of EMF-aware computation offloading has been presented in \cite{MerluzziEMF}, however in a SISO scenario without RISs.\\ 
\textbf{Our contribution.} The goal of this work is to explore the sinergy between RISs and MEC to enable computation offloading services, via \textit{blue communications} (with that referring to the human body specific absorption rate- SAR) \cite{Tesanovic2014}, exploiting both the benefits of RISs and MEC in future wireless networks.
To this end, we formulate the blue communication edge computing problem as a long-term optimization aiming to minimize the average EMFE, \textit{as per ICNIRP recommendations}, under MEC service delay constraints. We design an online algorithm able to dynamically configure RIS parameters, transmitter precoding, receiver combiner, and transmit power, with theoretical guarantees on system stability and asymptotic EMFE optimality. While precoding, combining, and RIS parameters are selected from generic codebooks, numerical results assess the performance with the use of accurate antenna patterns that take into account geometry, directivity, and array element patterns \cite{Clemente2012}. 
\\
\textit{Notation:} bold lower case and upper case letters denote vectors and matrices, respectively; the superscript $(\cdot)^T$ denotes the transposition operator; the operators $tr(\cdot)$ and $\textrm{diag}(\cdot)$ denote the trace, and the vectorization of the diagonal elements of a matrix; the operator $|\cdot|$ denotes the absolute value of a complex number, and $card(\cdot)$ is  the cardinality of a set; whereas, the long-term average of a random variable $X$ is denoted by $\overline{X}$ and defined as:
\begin{equation}\label{long_term_avg}
    \overline{X}=\lim_{T\to\infty}\frac{1}{T}\sum\nolimits_{t=1}^T\mathbb{E}\{X(t)\}
\end{equation}
\section{System Model}
\begin{figure}[htb!]
    \centering
   \includegraphics[width=.9\columnwidth]{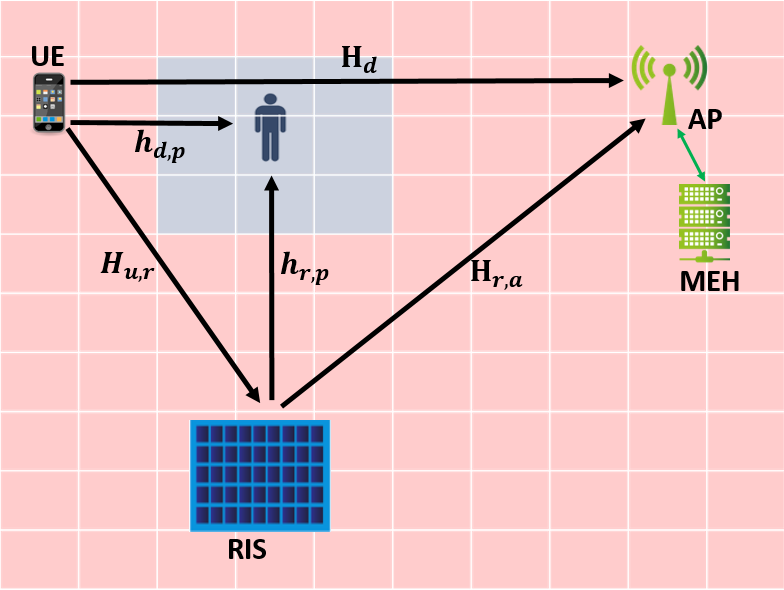}
    \caption{Network scenario}
    \label{fig:model}
\end{figure}
We consider the scenario of Fig. \ref{fig:model}, with a single user offloading computation tasks to a mobile edge host (MEH), through an RIS-aided wireless link with an access point (AP). From the EMFE perspective, several metrics have been proposed. However, ICNIRP recommendations consider the incident power density, averaged over $30$ minutes \cite{ICNIRP}. To this end, in this paper, we propose a long-term average optimization that perfectly matches this definition, rather than considering real-time measures that are not in line with existing recommendations. Thus, we consider time as organized in slots $t=1,2,\ldots$ of equal duration $\tau$. At the beginning of each slot, new offloading data are generated, new radio channels are observed and, based on these and other observations described later on, a new resource allocation decision is taken. In this way, while instantaneous instances of the EMFE may reach high values, the long-term average is finally minimized to achieve low exposure (as per the ICNIRP definitions and recommendations), under service delay guarantees. In the following, we describe the communication model, comprising the MIMO (direct and RIS-aided) communication channels, the precoding, combining and RIS response, the EMFE metric, and finally the computation model of the offloading service.
\subsection{Communication model}\label{sec:comm_model}
Let us consider a MIMO communication system, comprising a UE with an $N_u$ antenna array, an AP with an $N_a$ antenna array, and an RIS with $M$ elements. Then, the end-to-end communication channel at time $t$ can be described by a complex matrix $\mathbf{H}(t)\in\mathbb{C}^{N_a\times N_u}$. The latter entails two components, as depicted in Fig. \ref{fig:model}: i) a direct link $\mathbf{H}_{d}(t)\in \mathbb{C}^{N_a\times N_u}$ between the user and the AP; ii) an indirect link, comprising the channel $\mathbf{H}_{u,r}(t) \in \mathbb{C}^{{M}\times N_u}$ between the user and the RIS, and the channel $\mathbf{H}_{{r,a}}(t) \in \mathbb{C}^{N_a\times M}$ between the RIS and the AP. Finally, the E2E channel matrix can be written as follows \cite{channel}:
\begin{equation}\label{chan}
\mathbf{H}(t) = \mathbf{H}_{d}(t) +\mathbf{H}_{{r,a}}(t)\mathbf{\Theta}(t)  \mathbf{H}_{u,r}(t),
\end{equation}
where $\mathbf{\Theta}(t)\in\mathbb{C}^{M\times M}$ is a diagonal matrix whose $(i,i)$-th entry is $\mathbf{\Theta}_{i,i}(t)=e^{j\theta_i}$, with $\theta_i$ the phase shift of the $i$-th RIS's element. Also, we denote by $\mathbf{w}_u(t)\in\mathbb{C}^{N_u\times 1}$ the user precoding vector and, similarly, by $\mathbf{w}_a(t)\in\mathbb{C}^{N_a\times 1}$ the AP combining vector, both with normalized power, i.e. $tr(\mathbf{w}_u(t)\mathbf{w}^H_u(t))=1$ and $tr(\mathbf{w}_a(t)\mathbf{w}^H_a(t))=1$. The experienced data rate can then be written as follows:
\begin{equation}\label{uplink_rate}
          R_{u}(t) = W \log_2\left(1+\frac{\left|\mathbf{w}_{a}^H(t)\mathbf{H}(t)\mathbf{w}_{u}(t)\right|^2P_{\textrm{tx}}(t)}{N_0 W}\right)
\end{equation}
where $W$ represents the total communication bandwidth, $N_0$ is the noise power spectral density, and $P_{\textrm{tx}}(t)$ is the user transmit power at time $t$. Obviously, the data rate depends on the E2E channel states, as well as on the user precoding vector, the AP combining vector, the transmit power, and the RIS reflectivity matrix, which we will jointly optimize in the sequel. In this paper, we assume that the precoder, the combiner, and the RIS excitations can be selected from predefined codebooks that give rise to a set of candidate antenna patterns. The goal is to show that a naive choice of the antenna patterns (e.g. always transmitting towards the AP) is not necessarily the best choice in terms of EMFE-delay trade-offs, especially in dynamic time-varying environments. To this end, let us now formally characterized the EMFE and the E2E delay.
\subsection{Electromagnetic Field Exposure}
To evaluate EMF exposure, we consider space as divided in pixels $p\in\mathcal{P}$ of equal size, as in \cite{Chiaraviglio_densification,MerluzziEMF}, and as qualitatively represented in Fig. \ref{fig:model}. Then, assuming that humans possibly sojourn in one or more of these pixels (as for instance shown in the figure), our objective is to minimize a weighted sum of the EMFE in each pixel. In this way, if humans are not generally present in one pixel, the EMFE problem can be neglected in that particular location in space, assigning weight $0$. This allows us  to write a general optimization problem, to be customized on a case by case need. As an example, in the case of one human placed in pixel $p$, the objective becomes the $p$-th pixel EMFE, to achieve a blue communication area as depicted in Fig. \ref{fig:model}. This also represents the example investigated in the numerical results, while the overall method is more general and applies to multiple humans sojourning in multiple pixels. To this end, let us denote by $\mathbf{h}_{d,p}(t) \in \mathbb{C}^{1\times N_u}$ the direct channel vector between the UE and pixel $p$, and by $\mathbf{h}_{r,p}\in\mathbb{C}^{1\times M}$ the channel vector between the RIS and pixel $p$. Then, defining the overall channel between the user and pixel $p$ as $\mathbf{h}_p(t)=\mathbf{h}_{d,p}(t)+\mathbf{h}_{r,p}(t)\mathbf{\Theta}(t) \mathbf{H}_{u,r}(t)$, we can write the overall instantaneous power density as \cite{Chiaraviglio_densification}
\begin{equation}\label{emfe}
    P_{d,p}(t)=\frac{4\pi}{\lambda^2}P_{\textrm{tx}}(t)\left |\mathbf{h}_p(t)\mathbf{w}_u(t)\right|^2,
\end{equation}
where obviously both the direct and the reflected paths are taken into account. As already mentioned, our goal is to minimize a weighted sum of the average long-term EMFE in each pixel, which reads as follows (cf. \eqref{long_term_avg}):
\begin{equation}\label{weig_emfe}
    \overline{P_d}=\sum_{p\in\mathcal{P}}\omega_p\overline{P_{d,p}},\;\;p=1,\ldots,card(\mathcal{P})
\end{equation}
where $\omega_p$ are non negative weighting parameters. Obviously, minimizing the EMFE without any service constraints corresponds to not transmitting. Therefore, we now present the communication and computation queueing models and delays, to be kept bounded as part of a service quality requirement.
\subsection{Queuing Model and delay}
In our scenario, communication and computation phases take place in a subsequent way, as data are first buffered at the device before transmission and, once transmitted, they are buffered at the MEH before computation. We denote by $B_{l}(t)$ the communication buffer at the device, and by $B_{r}(t)$ the computation buffer at the MEH. Both buffers are measured in bits for the sake of simplicity, however different models considering data units (e.g. patterns to be classified at the MEH) can be envisioned under the same optimization framework \cite{Merluzzi_EML}. Their time evolution is described as follows.\\ 
\textbf{Communication buffer}: The local buffer welcomes all newly generated bits $A(t)$ at time slot $t$, while it is drained by transmitting them over the wireless interface at data rate $R_u(t)$ (cf. \eqref{uplink_rate}). Thus, its time evolution can be written as follows:
\begin{equation}\label{queue_evolution1}
    B_{l}(t+1)=\max\left(0,B_{l}(t)-\tau R_{u}(t)\right) +A(t)
\end{equation}
\textbf{Computation buffer}: Once data are uploaded, they are  buffered in a remote queue before computation. Therefore, the remote queue is fed by the data transmitted in uplink, and drained by the MEH processing the tasks. The time evolution of this queue can be written as follows:
\begin{align}\label{queue_evolution2}
\normalsize
  B_{r}(t+1)=&\max\left(0,B_{r}(t)-\tau f(t)/J\right)\nonumber\\    
              &+\min(B_{l}(t),\tau R_{u}(t)),
\end{align}
where $f(t)$ represents the amount of resources (in CPU cycles/s) allocated to each user during time slot $t$, and where we assumed a linear relation between the offloaded bits and the number of CPU cycles, i.e. $J$ is the number of CPU cycles per bit. Of course, more general relations between number of offloaded bits and CPU cycles may be envisioned, but are beyond the scope of this paper. In this work, we also assume that the CPU cycle frequency $f(t)$ is optimized, but it is rather selected at the edge server side on a per-slot basis, depending on its current availability. From the point of view of our orchestrator, $f(t)$ is a random variables of unknown statistics. Therefore, our method copes with the computation buffer through the uplink transmission, i.e. avoiding transmission when computation is congested at the MEH sever, as it will be clarified later on. Finally, due to the involvement of both buffers for the overall service, due to Little's law \cite{Little1961}, the end-to-end average service delay experienced by the device is directly linked to the average of the sum of both queues, and reads as  $\overline{D_u}=\tau\frac{\overline{B_{l}}+\overline{B_{r}}}{\overline{A}}$ (cf. \eqref{long_term_avg}). 

 \section{Problem formulation}
As already mentioned, our aim is to minimize the long-term average EMFE in selected areas across space (cf. \eqref{emfe}, \eqref{weig_emfe}) under service delay constraints, by dynamically and adaptively controlling the user precoding vector, the transmit power, the RIS reflectivity parameters, and the combining vector. We formulate the problem as follows:
 \begin{align}\label{lg_eq}
 \small
&\hspace{-1.5 cm }\min_{\{P_{\textrm{tx}}(t), \mathbf{w}_u(t), \mathbf{w}_a(t), \mathbf{\Theta}(t)\}_t} \quad  \overline{P_{d}}\\
\textrm{subject to} \;\;
&(a)\;  \overline{B_{l}} <\infty, \quad(b)\;  \overline{B_{r}} <\infty,\nonumber\\
    &(c)\;    0\leq P_{\textrm{tx}}(t)\leq P_{\textrm{tx}}^{\max},\, \forall t\quad(d)\; \mathbf{w}_u(t)\in\mathcal{W}_u,\, \forall t\nonumber\\
    &(e)\; \mathbf{w}_a(t)\in\mathcal{W}_a,\, \forall t\quad(f)\; \textrm{diag}\left\{\mathbf{\Theta}(t)\right\}\in\mathcal{T},\, \forall t.\nonumber
\end{align}  
where $\mathcal{W}_u$, $\mathcal{W}_a$, and $\mathcal{T}$ denote the precoder, the combiner, and the RIS codebooks. The constraints of \eqref{lg_eq} have the following meaning: $(a)$-$(b)$ the local and remote long-term average buffer lengths are limited, and so is the E2E service delay as a consequence; $(c)$ the transmit power is non negative and limited by a maximum value $P_{\textrm{tx}}^{\max}$; $(d)$-$(f)$ the precoder, the combiner, and the RIS parameters are chosen from their respective predefined sets, as anticipated in section \ref{sec:comm_model}. We will keep these codebooks generic, to then define them in the numerical results section, based on accurate antenna patterns that take into account the array element pattern, the geometry, and the directivity. Of course, the formulation in \eqref{lg_eq} and our proposed solution (described in the following) are not dependent on the codebooks, which can be defined and implemented according to the specific hardware constraints of different case studies. It should be noted that, despite the simplicity of the proposed single-user scenario, problem \eqref{lg_eq} is a priori very complex to solve, as it involves time averages over time-varying variables of unknown statistics, namely radio channels, arrivals, and computing availability. The proposed solution, described in the next section, allows us to decompose the problem and solve it in a per-slot basis, thus performing an exhaustive search over the codebooks in $(d)$-$(f)$, with corresponding closed form solutions for the transmit power. This is possible thanks to the definition of a suitable function to be minimized in each time slot, based only on instantaneous observations. Thanks to the theoretical foundations of Lyapunov stochastic optimization, this guarantees the queue stability constraints $(a)$-$(b)$, while asymptotically approaching the global optimal solution of \eqref{lg_eq} through one tuning parameter that trades off average EMFE optimality and service delay. Let us now formalize and elaborate on the above statements.
\subsection{Lyapunov stochastic optimization}
The goal of this section is to rigorously transform the original long-term problem into a per-slot problem, thus defining an instantaneous surrogate objective that, in the long-term, guarantees the desired performance. As it will be clarified later on, this new objective is a weighted sum of instantaneous EMFE and data rate, with the weights built on the buffer lengths and the aforementioned tuning parameter to be defined. In particular,  defining the vector
$\mathbf{b}(t)=[B_{l}(t),B_{r}(t)]$, let us first introduce the \textit{Lyapunov function} \cite{neely10}
\begin{equation}\label{lyap_fun}
    L(\mathbf{b}(t))=\frac{1}{2}\left[B_{l}^2(t)+B_{r}^2(t)\right],
\end{equation}
which is a measure of the overall buffer congestion state of the system. Our aim is to push the network towards low congestion states, while minimizing the weighted sum of EMFE. To this end, let us define the \textit{drift-plus-penalty} (DPP) function 
\begin{equation}\label{DPP}
   \Delta_{p}(t)=\mathbb{E}\{L(\mathbf{b}(t+1))-L(\mathbf{b}(t))+V\sum_{p\in\mathcal{P}}\omega_p P_{d,p}(t) |\mathbf{b}(t)\}, \nonumber
\end{equation}
where $P_{d,p}(t)$ is the instantaneous weighted sum of the incident power defined in \eqref{emfe} over all the pixels covering the area of interest. The DPP is the conditional expected change of the Lyapunov function over one slot, with a penalty factor, weighted by a parameter $V$, used to trade-off the instantaneous weighted sum power density and buffer backlogs. Theoretically speaking, buffers' stability (constraints $(a)$ and $(b)$ of \eqref{lg_eq}) is guaranteed if the DPP is bounded by a finite constant in each slot \cite{neely10}. We now proceed by minimizing a suitable upper bound of the DPP, as in \cite{neely10}. The upper bound, whose trivial derivations are reported in the following, reads as 
\begin{align}\label{DPP_UB}
    &\Delta_p(t)\leq C+ \mathbb{E}\{\left(B_{r}(t)-B_{l}(t)\right)\tau R_u(t)+A(t)B_{l}(t)\nonumber\\
    &-\tau B_{r}(t)f(t)/J+V\sum\nolimits_{p\in\mathcal{P}}\omega_p P_{d,p}(t)|\mathbf{b}(t)\}.
\end{align}
where $C$ is a positive finite constant that reads as
\begin{equation}\label{constant_C}
    C=\frac{1}{2}\left(A_{\max}^2+2(\tau R_u^{\max})^2+(\tau f_{\max})^2\right),
\end{equation}
with $A_{\max}$, $R_u^{\max}$, and $f_{\max}$ denoting the maximum number of arrivals, the maximum data rate, and the maximum CPU clock frequency, respectively, which are all finite by hypothesis.
\eqref{DPP_UB} and \eqref{constant_C} are obtained as follows. First, given a generic queue evolving as $Q(t+1)=\max(0,Q(t)-b(t))+A(t)$, we can write the following upper bound \cite{neely10}:
\begin{align}
    &\frac{Q^2(t+1)-Q^2(t)}{2}\leq \frac{A^2(t)+b^2(t)}{2}-A(t)\min(b(t),Q(t))\nonumber\\
    &+Q(t)(A(t)-b(t))\leq \frac{A_{\max}^2 + b_{\max}^2}{2} + Q(t)(A(t)-b(t)),\nonumber
\end{align}
where $A_{\max}$ and $b_{\max}$ are finite upper bounds of $A(t)$, $b(t)$, which exist by hypothesis, as already stated. Then, recalling \eqref{queue_evolution1} and \eqref{queue_evolution2}, and applying the above upper bound to \eqref{DPP}, we easily obtain \eqref{DPP_UB} and \eqref{constant_C}.
Now, by greedily minimizing \eqref{DPP_UB} in each time slot (i.e. removing the expectation), we obtain the following per-slot problem, which exploits the desired deterministic instantaneous objective function to be used for the dynamic and adaptive selection of precoding, decoding, RIS parameters, and transmit power (we omit the temporal index to lean the notation):
 \begin{align}\label{slot_problem_radio}\small
     \min_{{P_{\textrm{tx}}}, \mathbf w_u, \mathbf w_a, \mathbf{\Theta}} \;  &V\sum\nolimits_{p\in\mathcal{P}}\omega_p P_{d,p}+(B_r-B_l)R_u\\
   &\textrm{subject to} \quad (c)\textrm{-}(f)\;\textrm{of}\;\eqref{lg_eq},\nonumber
\end{align}
where we recall that $P_{d,p}$ (cf. \eqref{emfe}) and $R_u$ (cf. \eqref{uplink_rate}) are functions of precoding, combining, RIS parameters, and transmit power. As a direct consequence of \cite[th. $4.8$]{neely10}, by solving \eqref{slot_problem_radio} in each slot, buffers' stability is guaranteed, with an upper bound on their average length (i.e. E2E delay) that grows as $O(V)$, with $V$ defined in \eqref{DPP}.  Moreover, the distance from the global optimal solution of the original problem \eqref{lg_eq} and the one obtained through \eqref{slot_problem_radio}, decreases as $O(1/V)$. 
More practically, for $V$ finite, a corresponding E2E delay and EMFE are obtained, i.e. a performance trade-off is explored by varying this single tuning parameter, as anticipated in previous section. The last effort is to solve \eqref{slot_problem_radio}. To this end, let us notice that, in practical implementations, it is not likely to have high cardinality sets for $\mathcal{W}_u$, $\mathcal{W}_a$, and $\mathcal{T}$, so that an exhaustive search over the feasible set of \eqref{lg_eq} is possible, of course with the exception of the transmit power, which is a continuous variable. Nevertheless, given $\mathcal{W}_u$, $\mathcal{W}_a$, and $\mathcal{T}$, the transmit power can be found in closed form. In particular, it is straightforward to observe that, whenever $B_r\geq B_l$, the optimal solution is to not transmit (i.e. $P_\textrm{tx}=0$), as both terms in \eqref{slot_problem_radio} are monotonic non decreasing functions of $P_\textrm{tx}$. On the other hand, whenever $B_r< B_l$, the problem is convex, and the optimal $P_{\textrm{tx}}$ can be found in closed form through the Karush-Kuhn-Tucker conditions as follows (simple derivations are omitted due to the lack of space) \cite{Boyd2004}:
\begin{equation}\label{opt_power}
    P_{\textrm{tx}}\!=\!\!\left[\frac{W(B_l-B_r)\lambda^2}{\displaystyle4\pi V\ln(2)\sum_{p\in\mathcal{P}}\omega_p\left |\mathbf{h}_p\mathbf{w}_u\right|^2}\!-\!\frac{\left|\mathbf{w}_{a}^H\mathbf{H}\mathbf{w}_{u}\right|^2}{N_0 W}\right]_0^{P_{\textrm{tx}}^{\max}}
\end{equation}
Then, the global optimal solution of \eqref{slot_problem_radio} is simply obtained by computing $P_{\textrm{tx}}$ for all possible precoders, combiners, and RIS configurations, and selecting the solution that achieves the lowest value of the objective function in \eqref{slot_problem_radio}. Overall, in each slot, one needs to observe queue states, wireless channels, and data arrivals, solve \eqref{slot_problem_radio} through the above mentioned procedure, and finally update the buffers accordingly. Surprisingly, starting from a long-term complex optimization problem, we end up to a deterministic one that can be solved through exhaustive searches over limited sets and closed form solutions, thanks to the definition of an instantaneous objective that weights service queue states and EMFE.
\section{Numerical Results}
\begin{figure}[t!]
    \centering
   \includegraphics[width=.9\columnwidth]{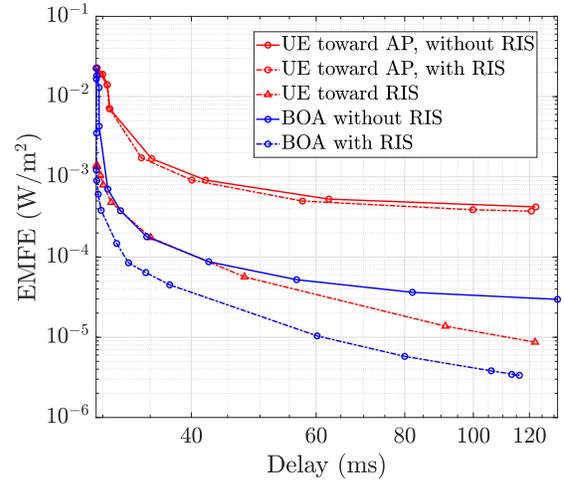}
    \caption{EMFE-Delay trade-off}
    \label{fig:boa_delay}
\vspace{-.4 cm}
\end{figure}

\begin{figure}[t!]
    \centering

   \includegraphics[width=.9\columnwidth]{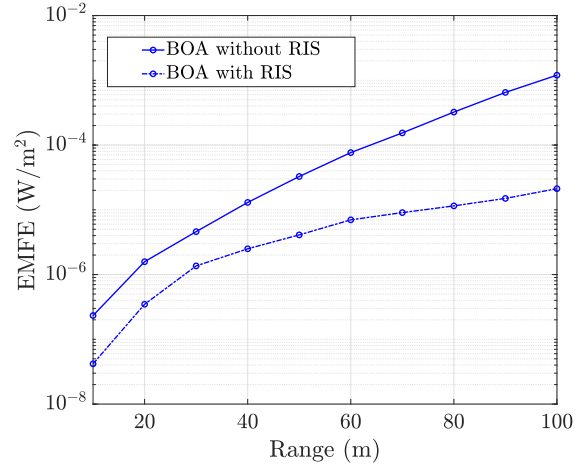}
    \caption{EMFE-Coverage trade-off}
    \label{fig:boa_range}
\vspace{-.6 cm}
\end{figure}

In this section, numerical results are provided to assess the impact of our proposed optimization algorithm on the EMFE-delay trade-off. We consider a scenario with a UE aiming to offload his tasks to a MEH collocated at his serving AP. The user is assigned a  bandwidth of $B=800$ MHz, while the noise power spectral density is set to $N_0=-174$ dBm/Hz. The slot duration is set to $\tau=10$ ms while for the user, the arrival rate is set to $10$ Gbps with Poisson distribution. 
At each time slot,  all channels (cf. \eqref{chan}) are generated using a Rician model, for a typical mmWave operating frequency, $f=28$ GHz, with $N_u=8$, $N_a=8$, and $M=20$, with element spacing $\lambda/2$. The antenna patterns used to build $\mathcal{W}_u$, $\mathcal{W}_a$, and $\mathcal{T}$ are taken from \cite{Haut10}, with each element modeled as in \cite[Eqn. $2$]{Clemente2012}. A range of $-60^{\circ}$ to $60^{\circ}$, with a step of $10^{\circ}$ is considered for the UE and the AP, while a range of $-30^{\circ}$ to $30^{\circ}$ with a step of $5^{\circ}$ is considered for the RIS, with $0^{\circ}$ the direction perpendicular to the array. The maximum transmit power for the  user  is set to $P_{tx}^{\max}=100$ mW. At the MEC side, we assume the MEH to be able to accomodate all requests on average, however with an instantaneous random $f(t)$ uniformly distributed.
In particular, denoting by (x, y, z) the 3D coordinates of an element, we model the scenario deployment depicted in Fig. \ref{fig:model} using the following positions: the AP at (50,50,1), the UE at (0,50,1), the RIS at (4,48,1), and the man at (1,50,1). Note that, for all elements, we use the uniform linear array (ULA) case. \\
As a first result, in Fig. \ref{fig:boa_delay}, we show  the trade-off between the EMFE and the average E2E delay, obtained with our method by tuning the trade-off parameter V. For this simulation, we consider four different benchmark comparisons: i) the no RIS-aided case, with the UE always transmitting towards the AP, and with transmit power optimized as in \eqref{opt_power}; ii) the RIS-aided case, with the UE always transmitting towards the AP, and with transmit power optimized as in \eqref{opt_power}; iii) the RIS-aided case, with the UE always transmitting towards the RIS, and with transmit power optimized as in \eqref{opt_power}; iv) the case without the RIS, but applying our optimization method. Finally, we term our full optimization algorithm as BOA (blue optimization algorithm).
In Fig. \ref{fig:boa_delay}, results are obtained by increasing the Lyapunov trade-off parameter V from right to  left. For all curves we can notice how, by increasing V, the system average EMFE decreases while the average service delay increases. Results show how  the use of the RIS offers the opportunity to reduce the level of EMFE for a given service delay, also in the case the UE always transmits towards the RIS. This is due to the fact that the pixels of interest are between the UE and the AP in this simulation. However, this gain is considerably enhanced when BOA is applied, due to the increased degrees of freedom introduced by the adaptive selection of precoding, combining, and RIS parameters. In other words, thanks to the fact that the EMFE is defined as an average metric, the user can opportunistically exploit the high data rate direct link from time to time to upload a considerable amount of data, thus increasing the instantaneous EMFE, provided that the indirect link is exploited as backup to lower the long-term average. The gain of the RIS-aided communication channel is indeed linked to this. Hereafter, to point out more the impact of the RIS and the BOA, we illustrate, through Fig. \ref{fig:boa_range}, the average EMFE as a function of the transmission's range (i.e. the distance between the UE and the AP), for a fixed E2E delay bound of 100 ms, obtained by tuning the trade-off parameter V. 
\begin{figure}[htb!]
    \centering

   \includegraphics[width=.9\columnwidth]{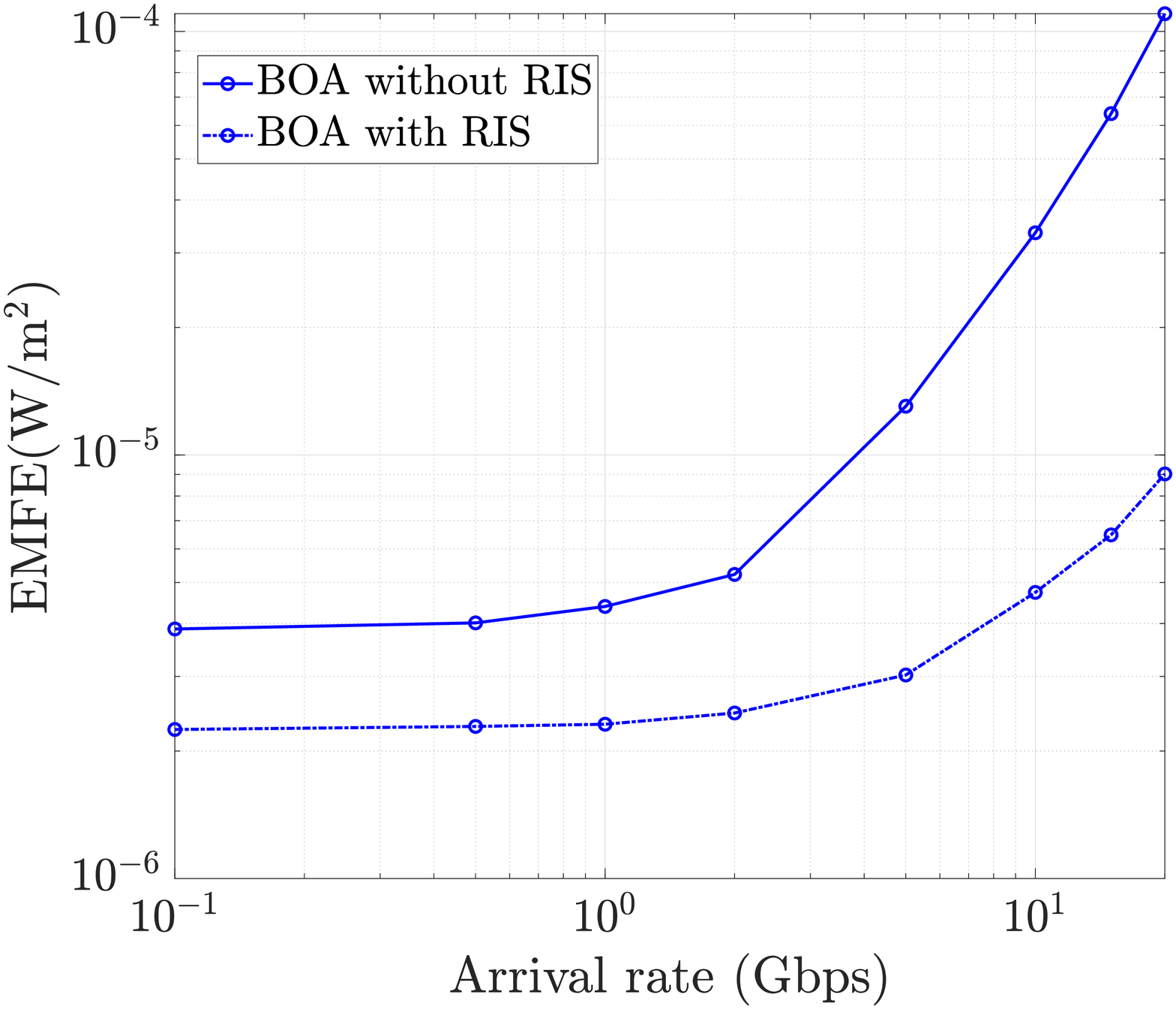}
    \caption{EMFE-Rate trade-off}
    \label{fig:boa_rate}
    \vspace{-.6 cm}
\end{figure}
It can be clearly noticed how for both cases, the EMFE level increases with  the range toward the AP, as expected, as a higher transmit power is needed at the UE side to reach the AP with enough signal quality. However, we can notice how the use of the RIS helps extending the communication range, with less impact on the exposure, with arounf $15$ dB of gain at $100$ m.\\
As a final result, we illustrate how the proposed algorithm behaves with respect to the MEC traffic arrival rate, i.e the necessary data rate over the wireless interface to guarantee stability. 
As shown from Fig. \ref{fig:boa_rate}, the case of the BOA with RIS yields considerable gains for different arrival rates. More specifically, for  rates ranging from $100$ Mbps to  $1$ Gbps, the gap is stable around $3$ dB, while it increases to more than $10$ dB for an arrival rate of $20$ Gbps. This trivially suggests that the RIS is more beneficial especially for high data rate services, as for low data rate services the UE can keep low power communications while still guaranteeing stability and, as a consequence, low average EMFE.

\section{Conclusion}
We proposed an online method able to adaptively and jointly optimize precoding, combining, RIS parameters, and transmit power in a RIS-aided MEC offloading scenario. As objective, we considered the average EMFE in selected areas within the service coverage, with constraints on the E2E service delay. 
We reduced a long-term problem to a per-slot optimization, which allowed us to solve it through a low complexity procedure involving an exhaustive search over low cardinality sets, coupled with a closed form solution for the transmit power. Numerical results show the effectiveness of our method and the benefits of the RIS in enabling blue communications for computation offloading services, due to which the uplink direction of communication will be exploding in future 6G systems. Future directions include multi-user scenarios, but also practical phase shifts and antenna losses at UEs, APs and RISs, to investigate the impact of hardware constraints on the results shown in the paper. However, such practical constraints will rather impact absolute performance, while we expect them to provide similar relative gains.

\end{document}